\documentclass[aps,superscriptaddress,showpacs,preprint]{revtex4}
\usepackage{amsfonts}
\usepackage{amsmath}
\usepackage{amssymb}
\usepackage{graphicx}
\usepackage[T1]{fontenc}
\usepackage[latin9]{inputenc}
\usepackage{array}
\usepackage{longtable}
\usepackage{amsmath}
\usepackage{esint}

\begin{document}
\title{Quantum molecular dynamics simulations of the thermophysical properties
of shocked liquid ammonia for pressures up to 1.3 TPa}
\author{Dafang Li}
\affiliation{LCP, Institute of Applied Physics and Computational
Mathematics, Beijing 100088, People's Republic of China}
\affiliation{Data center for high energy density physics, Institute
of Applied Physics and Computational Mathematics, Beijing 100088,
People's Republic of China}
\author{Ping Zhang}
\thanks{zhang\_ping@iapcm.ac.cn}
\affiliation{LCP, Institute of Applied Physics and Computational
Mathematics, Beijing 100088, People's Republic of China}
\affiliation{Center for Applied Physics and Technology, Peking
University, Beijing 100871, People's Republic of China}
\author{Jun Yan}
\thanks{yan\_jun@iapcm.ac.cn}
\affiliation{LCP, Institute of Applied Physics and Computational
Mathematics, Beijing 100088, People's Republic of China}
\affiliation{Data center for high energy density physics, Institute
of Applied Physics and Computational Mathematics, Beijing 100088,
People's Republic of China}
\affiliation{Center for Applied Physics
and Technology, Peking University, Beijing 100871, People's Republic
of China}


\begin{abstract}
We investigate via quantum molecular-dynamics simulations the
thermophysical properties of shocked liquid ammonia up to the
pressure 1.3 TPa and temperature 120000 K. The principal Hugoniot is
predicted from wide-range equation of state, which agrees well with
available experimental measurements up to 64 GPa. Our systematic
study of the structural properties demonstrates that liquid ammonia
undergoes a gradual phase transition along the Hugoniot. At about
4800 K, the system transforms into a metallic, complex mixture state
consisting of $\textnormal{N}\textnormal{H}_{3}$,
$\textnormal{N}_{2}$, $\textnormal{H}_{2}$, N, and H. Furthermore,
we discuss the implications for the interiors of Uranus and Neptune.
\end{abstract}
\maketitle

\section{INTRODUCTION}

Ammonia, together with water and methane, are the major constituents
of the giant planets in our solar system. In particular, Uranus and
Neptune are thought to have \textquotedblleft hot
ice\textquotedblright\ layers predominantly made up of 56\% water,
36\% methane, and 8\% ammonia in proportions
\cite{Hubbard1980,Stevenson1982}. The behavior of these molecular
compounds at extreme conditions (temperatures $T>2000$ K, pressures
$P>20$ GPa) is therefore crucial for understanding Uranus and
Neptune physics, including gravitational moments, atmospheric
composition, and magnetic field \cite{Stevenson1983}. As a weakly
hydrogen-bonded liquid, the thermophysical properties of ammonia at
high pressure and high temperature are of fundamental interest both
for astrophysics and solid state physics.

Experimental researches on ammonia are indispensable for exploring
its characteristics under extreme conditions. Standard explosive
techniques were firstly used to get the Hugoniot curve of ammonia up
to 39 GPa. In this pressure region, no transition was observed along
the Hugoniot \cite{Dick1981}. Mitchell \textit{et al}. have used
two-stage light-gas gun to reach pressure of 64 GPa
\cite{Mitchell1982}. The first shock temperature measurements on
ammonia were performed at pressures of 61 and 48 GPa for diagnosing
the physics occurring at these extreme conditions
\cite{Radousky1990}. Through comparing with the calculations using
fluid perturbation theory \cite{Radousky1986}, it was concluded that
there may exist absorption mechanism at these high pressures.
Previous electrical conductivity measurements have shown that
ammonia becomes conductive in the pressure range of 7-28 GPa
\cite{Kovel1973,Nellis1988}, which is thought to be induced by
molecular dissociation-ionization. Several static experimental
studies performed in diamond anvil cells were limited to the phase
diagram up to 60 GPa and 2500 K \cite{Ninet2008,Li2009,Ojwang2012}.

On theoretical side, ammonia has been predicted to become a protonic
conductor above 60 GPa and 1200 K using \textit{ab} \textit{initio}
molecular dynamics \cite{Cavazzoni1999}. At even more extreme
conditions, ammonia molecules are expected to dissociate and react
at very rapid rates \cite{Cavazzoni1999,Chau2011}. In addition,
first-principles calculations have shown that driven by the entropy
of mixing term in the free energy formulation \cite{Ojwang2012},
ammonia molecules chemically dissociate to $\textnormal{N}_{2}$ and
$\textnormal{H}_{2}$ above approximately 7 GPa and 900 K.

In this paper, we perform the first comprehensive quantum molecular
dynamics (QMD) simulations of the high-pressure and high-temperature
behavior of ammonia with densities and temperatures ranging from 0.7
g/cm$^{3}$ and 230 K to 2.6 g/cm$^{3}$ and 120000 K along the
principal Hugoniot. We determine the equation of state (EOS) in the
warm dense region by means of QMD, where the active electrons are
treated in a full quantum mechanical way within the
finite-temperature density functional theory (FT-DFT). This method
has been proven to be a successful tool to calculate physical
properties of complex plasmas under such extreme conditions
\cite{Collins2001,Mazevet2003,Kress2001,Desjarlais2002,Mazevet2004,Laudernet2004}.
In combination with the Kubo-Greenwood formulation, we derive the
electrical and optical properties in order to locate the
nonmetal-to-metal transition. Through analyzing the concentration of
molecular species along the Hugoniot based on pair correlation
function, we explore and describe the mechanism for this transition.

\section{COMPUTATIONAL METHOD}

\subsection{Quantum molecular dynamics}

The particular implementation of QMD method used in the present
study comes from the Vienna ab initio simulation package (VASP)
plane-wave pseudopotential code developed at the Technical
University of Vienna \cite{Kresse1993,Kresse1996}, in the framework
of a FT-DFT \cite{Lenosky2000,Bagnier2001}. The electronic states
are populated according to the Fermi-Dirac statistics, with
electronic temperature set equal to that of ions. We consider the
electronic states occupied down to
$\textnormal{1\ensuremath{0^{-6}}}$. The electron wavefunctions are
calculated using the all-electron projector augmented wave (PAW)
potentials \cite{Kresse1999,Bloch1994}. The Perdew-Wang 91
parametrization of the generalized gradient approximation (GGA)
\cite{Perdew1991} is employed for the exchange-correlation energy.
Atoms move classically according to the forces, which originate from
the interactions of ions and electrons.

QMD simulations are performed in the canonical (NVT) ensemble with
Nose-Hoover thermostat for selected densities from 0.7 to 2.6
g/cm$^{3}$ and temperatures from 230 to 120000 K that highlight the
single-shock Hugoniot region. 27 nitrogen and 81 hydrogen atoms
(twenty-seven ammonia molecules) are treated in a cubic cell of the
size appropriate to the considered density. We fix the plane-wave
cutoff at 550.0 eV which is tested to give good convergence for both
total energy and pressure. The Brillouin zone sampling of 108-atom
calculations use only the $\Gamma$ point for molecular dynamics,
while $4\times4\times4$ Monkhorst-Pack \cite{Monkhorst1976} scheme
$k$ points for the calculations of electronic properties.
Integration of the equations of motion proceed with time step of
0.5-1.0 fs for different pressure-temperature ranges. After about 3
ps the system is equilibrated and the subsequent 5 ps are taken to
calculate the EOS and electronic properties as running averages. The
ion temperature $T_{i}$ is fixed using velocity scaling, while the
electron temperature $T_{e}$ is in turn set to that of the ions
$T_{i}$ based on the assumption of local thermodynamical
equilibrium.

\subsection{Optical properties}

At the Hugoniot points the electronic properties are calculated for
ten configurations selected from an equilibrated (in an average
sense) portion of the molecular dynamics run. The configurations are
spaced at time steps separated by at least the correlation time, the
$e-$folding time of the velocity autocorrelation function. For each
of these configurations, the Kubo-Greenwood formulation is used to
calculate the electrical conductivity, without particular
assumptions made on the ionic structure or on the electron-ion
interactions. In the framework of the quasi-independent particle
approximation, the Kubo-Greenwood formulation
\cite{Kubo1957,Greenwood1958} gives the real part of the electrical
conductivity as a function of frequency $\omega$,
\begin{eqnarray}
\sigma_{1}\left(\omega\right) & = & \frac{2\pi}{3\omega\Omega}\underset{\mathbf{k}}{\sum}w\left(\mathbf{k}\right)\overset{N}{\underset{j=1}{\sum}}\overset{N}{\underset{i=1}{\sum}}\overset{3}{\underset{\alpha=1}{\sum}}\left[f\left(\epsilon_{i},\mathbf{k}\right)-f\left(\epsilon_{j},\mathbf{k}\right)\right]\nonumber \\
 &  & \times\left|\left\langle \Psi_{j,\mathbf{k}}\right|\nabla_{\alpha}\left|\Psi_{i,\mathbf{k}}\right\rangle \right|^{2}\delta\left(\epsilon_{j,\mathbf{k}}-\epsilon_{i,\mathbf{k}}-\hbar\omega\right), \end{eqnarray}
where $\omega$ is the frequency, $\Omega$ is the atomic volume, and
$N$ is the total number of bands used. $\Psi_{i,\mathbf{k}}$ and
$\epsilon_{i,\mathbf{k}}$ are the electronic eigenstate and
eigenvalue for the electronic state $i$ at $\mathbf{k}$,
$f\left(\epsilon_{i},\mathbf{k}\right)$ stands for the Fermi
distribution function, and $w\left(\mathbf{k}\right)$ represents the
$\mathbf{k}-$point weighting factor. Other properties can be
directly derived from the frequency-dependent real part of the
electrical conductivity. The imaginary part
$\sigma_{2}\left(\omega\right)$ is obtained by using the
Kramer-Krönig relation

\begin{eqnarray}
\sigma_{2}\left(\omega\right) & = &
-\frac{2}{\pi}P\int\frac{\sigma_{1}\left(\nu\right)\omega}{\left(\nu^{2}-\omega^{2}\right)}d\nu,\end{eqnarray}
where $P$ denotes the principal value of the integral. The complex
dielectric function, in turn, follows immediately from the complex
conductivity,

\begin{eqnarray}
\epsilon_{1}\left(\omega\right) & = & 1-\frac{4\pi}{\omega}\sigma_{2}\left(\omega\right),\end{eqnarray}

\begin{eqnarray}
\epsilon_{2}\left(\omega\right) & = &
\frac{4\pi}{\omega}\sigma_{1}\left(\omega\right).\end{eqnarray} And
then the real $n\left(\omega\right)$ and imaginary
$k\left(\omega\right)$ parts of the index of refraction have a
relation with the complex dielectric function by a simple formula,

\begin{eqnarray}
\epsilon\left(\omega\right) & = &
\epsilon_{1}\left(\omega\right)+i\epsilon_{2}\left(\omega\right)=\left[n\left(\omega\right)+ik\left(\omega\right)\right]^{2}.\end{eqnarray}
Finally, the reflectivity $r\left(\omega\right)$ and absorption
coefficient $\alpha\left(\omega\right)$ can be determined from these
quantities as follows

\begin{eqnarray}
r\left(\omega\right) & = & \frac{\left[1-n\left(\omega\right)\right]^{2}+k\left(\omega\right)^{2}}{\left[1+n\left(\omega\right)\right]^{2}+k\left(\omega\right)^{2}},\end{eqnarray}

\begin{eqnarray}
\alpha\left(\omega\right) & = & \frac{4\pi}{n\left(\omega\right)c}\sigma_{1}\left(\omega\right).\end{eqnarray}

\section{Results and Discussions}

\subsection{Hugoniot}

\begin{table}
\caption{Points along the principal ammonia Hugoniot derived from
DFT-MD simulations at a series of density $\left(\rho\right)$,
pressure $\left(P\right)$, and temperature $\left(T\right)$.}
\begin{center}
\begin{tabular}{ccccccccc}
\hline \hline
$\rho$ & & $P$ & & $T$ & & $U_{p}$ & & $U_{s}$\\
$(\textnormal{g/cm\ensuremath{^{3}}})$ & & (GPa) & & (K) & & (Km/S) & & (Km/S)\\
\hline 0.80 & & 1.24 & & 250 & & 0.487 & & 3.655\\
1.00 & & 3.73 & & 308 & & 1.284 & & 4.186\\
1.2 & & 14.33 & & 1735 & & 2.955 & & 6.998\\
1.4 & & 25.41 & & 2295 & & 4.301 & & 8.521\\
1.6 & & 41.43 & & 3082 & & 5.820 & & 10.269\\
1.7 & & 51.53 & & 3562 & & 6.634 & & 11.203\\
1.8 & & 63.28 & & 4174 & & 7.491 & & 12.184\\
1.9 & & 77.09 & & 4791 & & 8.404 & & 13.232\\
2.0 & & 95.94 & & 6036 & & 9.509 & & 14.554\\
2.2 & & 221.44 & & 19180 & & 14.790 & & 21.595\\
2.4 & & 567.67 & & 54229 & & 24.130 & & 33.932\\
2.6 & & 1273.65 & & 112663 & & 36.705 & & 50.051\\
\hline\hline
\end{tabular}
\end{center}
\end{table}

As a crucial measure for theoretical EOS data, the Hugoniot
describes the locus of states satisfying the Rankine-Hugoniot
equation \cite{Zeldovich1966}, which follows from the conservation
of mass momentum and energy across the shock front. The initial and
final internal energy, pressure and volume, respectively,
$\left(E_{0},\: P_{0},\: V_{0}\right)$ and $\left(E,\: P,\:
V\right)$ are related through such an equation in the form
\begin{eqnarray}
\left(E_{1}-E_{0}\right)+\frac{1}{2}\left(V_{0}-V_{1}\right)\left(P_{0}+P_{1}\right)
& = & 0,\end{eqnarray} where the internal energy $E$ equals to the
sum of the ion kinetic energy $\frac{3}{2}k_{B}T_{i}$, the time
average of the DFT potential energy, and zero-point energy. The
pressure consists of contributions from the electronic $P_{e}$ and
ionic $P_{i}$ components, which come from, respectively, the
derivatives taken with respect to the Kohn-Sham electronic orbitals
and the ideal gas expression since ions move classically. We thus
have $P=P_{e}+\rho_{n}k_{B}T$, where $\rho_{n}$ is the number
density. The particle velocity of material behind the shock front
$u_{p}$ and the shock velocity $u_{s}$ are then determined from the
other two Rankine-Hugoniot equations \cite{Zeldovich1966},
\begin{eqnarray}
V_{1} & = &
V_{0}\left[1-\left(u_{p}/u_{s}\right)\right],\end{eqnarray}
\begin{eqnarray}
P_{1}-P_{0} & = & \rho_{0}u_{s}u_{p}.\end{eqnarray} In the present
work, the initial internal energy of ammonia is calculated to be
$E_{0}=-18.64$ eV/molecule under the experimental condition with
$\rho_{0}=0.6933$ g/cm$^{3}$ at a temperature of 230 K. The initial
pressure can be neglected when compared to the high pressure of
shocked state along the Hugoniot. To find the Hugoniot points for a
given $V_{1}$, a series of simulations are performed for different
temperatures $T$. $E_{1}$ and $P_{1}$ are then fitted to a cubic
function of $T$. We plot $\left(E_{1}-E_{0}\right)$ and
$\frac{1}{2}\left(V_{0}-V_{1}\right)\left(P_{0}+P_{1}\right)$ as a
function of $T$, and the intersection fixes the point satisfying Eq.
(8). The principal Hugoniot points of ammonia derived from
Rankine-Hugoniot equations are listed in Table I.

\begin{figure}
\includegraphics[clip,scale=0.5]{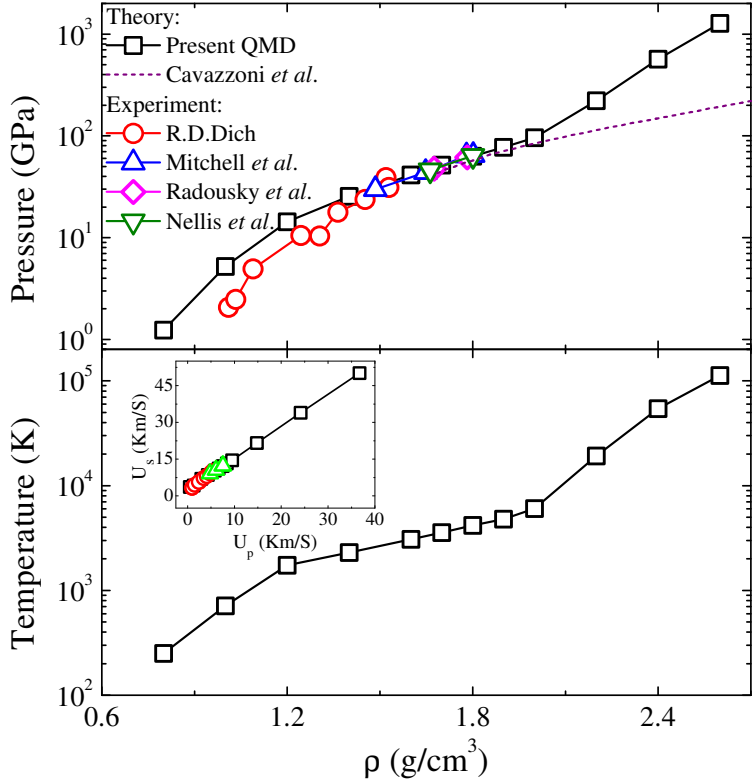}
\caption{Principal Hugoniot of liquid ammonia. For comparison,
previous experimental data (Ref. 4, Ref. 5, Ref. 6 and Ref. 9) and
theoretical results along planetary isentrope (Ref. 13) are also
included.}
\end{figure}

Figure 1 shows the pressure and temperature as functions of density
along the Hugoniot curve for ammonia, along with experimental
measurements and results from previous \textit{ab} \textit{initio}
molecular dynamic simulations along the planetary isentrope for
comparison. We find a very good agreement with the experimental
results all along the single-shock Hugoniot up to 64 GPa, while we
also predict the behavior of ammonia under higher pressures and
temperatures, which may be testified in future experiments. The data
along the planetary isentrop of Ref. 13 presents a prominent
discrepancy with our QMD results, which is because that the
temperature along the planetary isentrope is different from the
shock temperature. In addition, the Hugoniot curve is naturally
divided into three segments with their respective features, similar
to hot dense methane \cite{Sherman2012}. The temperature is found to
increase linearly with density up to 2000 K, which results from the
fact that ammonia remains its ideal molecular configurations without
dissociation. In the temperature range of 2000-6000 K, a plateau
appears and the temperature no longer increases as rapidly as the
corresponding density. As discussed in more detail later, this
region coincides with the onset of a significant fraction of
molecular dissociation and the transformation to a complex mixture
consisting of a variety of species, including NH$_{3}$, H$_{2}$,
N$_{2}$, H, and N. Furthermore, it will be reported that ammonia
becomes metallic at these temperature-pressure conditions. Beyond
6000 K the temperature increases rapidly with density again with the
slop depending on the initial density. In this regime, some ionic
species exist with very short lifetime and the system enters into a
plasma state.

The Hugoniot curves for several other initial densities and
temperatures are also presented in Fig. 2 in order to explore
different pressure-temperature conditions of interest to future
experimental measurements. It can be seen that subtle changes in the
starting density allow liquid ammonia to reach different regions of
the pressure and temperature, while the moderate variations in the
initial temperature do not. It is highly desirable as a means for
one to understand the properties of liquid ammonia over a wide range
of extreme conditions. These predicted Hugoniots may be verified by
utilizing modern experimental approaches.

\begin{figure}
\includegraphics[clip,scale=0.5]{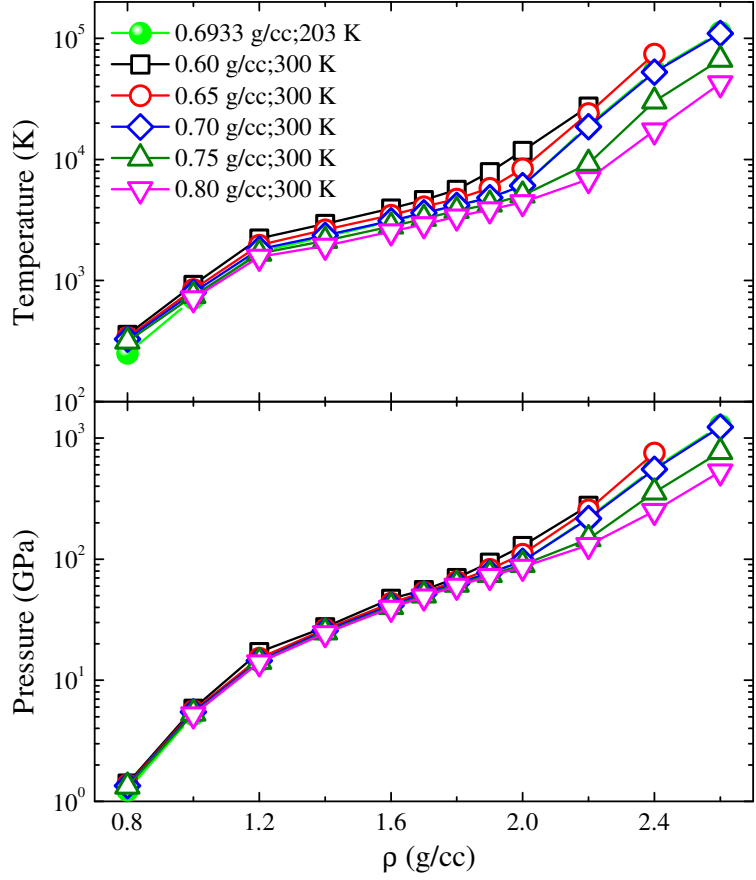}
\caption{$P-\rho$ and $T-\rho$ Hugoniots computed for several
initial conditions.}
\end{figure}

We further plot temperature as a function of pressure along each of
the Hugoniot curves with different initial densities in Fig. 3, in
which the isentropes for Uranus and Neptune are also included
\cite{Redmer2011}. At approximately 2000-6000 K, with ammonia
entering into complex mixture state, the slopes change obviously. As
can be seen from Fig. 3, the isentropes intersect these Hugoniot
curves just in this region. It implies that ammonia does not remain
its ideal molecular form in the interiors of Uranus and Neptune, but
transforms into a mixture of NH$_{3}$, H$_{2}$, N$_{2}$, H, and N.
In addition, the phase diagram of ammonia determined by Cavazzoni
\textit{et al}. \cite{Cavazzoni1999} has shown that a superionic
phase exists far below the isentropes of Uranus and Neptune. And
thus it could be predicted that molecular hydrogen may be expelled
from the interiors of Uranus and Neptune into the outer layer.  The
cores of these planets may become more compact.

\begin{figure}
\includegraphics[clip,scale=0.5]{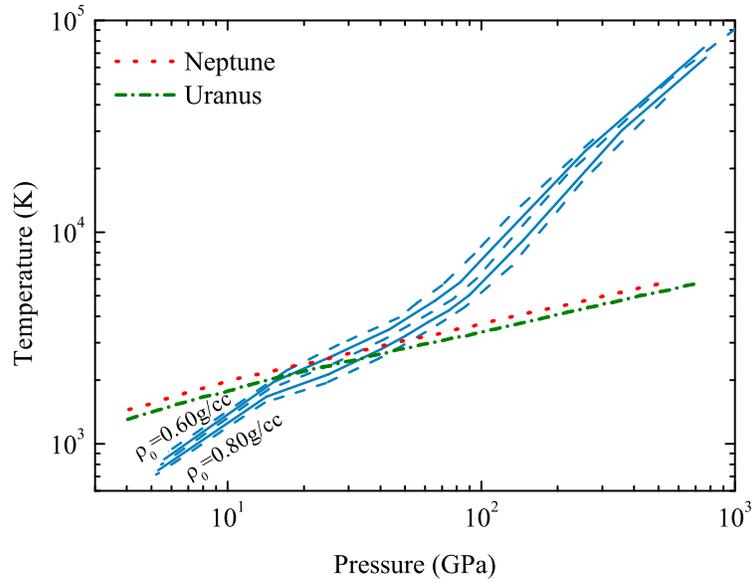}
\caption{Temperature as a function of pressure along the shock
Hugoniot curves from Fig. 2 for different initial densities. The
isentropes for Uranus and Neptune are also included.}
\end{figure}
\subsection{Liquid structure}

To quantify the structural change in ammonia along the Hugoniot, we
calculate the pair-correlation function for each possible species of
NH$_{3}$, N$_{2}$, and H$_{2}$. The pair-correlation function gives
the possibility of finding an atom of a given type at a given
distance from a reference atom. The results are presented in Fig. 4.
At the lowest density $\rho\mathtt{=}$0.6933 g/cm$^{3}$, the sharp
peak at about $1.02 \ \textrm{\AA}$\ corresponding to the
equilibrium internuclear distance of the N-H bond in ammonia
molecule and the following minimum close to zero indicate that
ammonia remains its ideal molecular configurations stably. As the
density is raised to $\rho\mathtt{=}$1.6 g/cm$^{3}$, new
$\textnormal{N-N}$ peak begins to emerge at the equilibrium distance
of nitrogen molecule 1.2 $\textrm{\AA}$. Meanwhile, the maxima of
$g_{\textnormal{N-H}}\left(r\right)$ are reduced and broadened
significantly. It is indicated that ammonia molecules dissociate and
small amount of nitrogen molecules form at this density. At 1.8
g/cm$^{3}$, N-H peak continues to be reduced and broadened while the
peaks of N-N and H-H increase. This trend persists up to a density
of 2.0 g/cm$^{3}$. With increasing the density further, the N-H peak
diminishes, while N-N and H-H peaks are  broadened and even
flattened. This suggests that all molecules are very short-lived and
unstable at these extreme conditions .
\begin{figure*}
\includegraphics[bb=0bp 0bp 568bp 292bp,clip,scale=0.7]{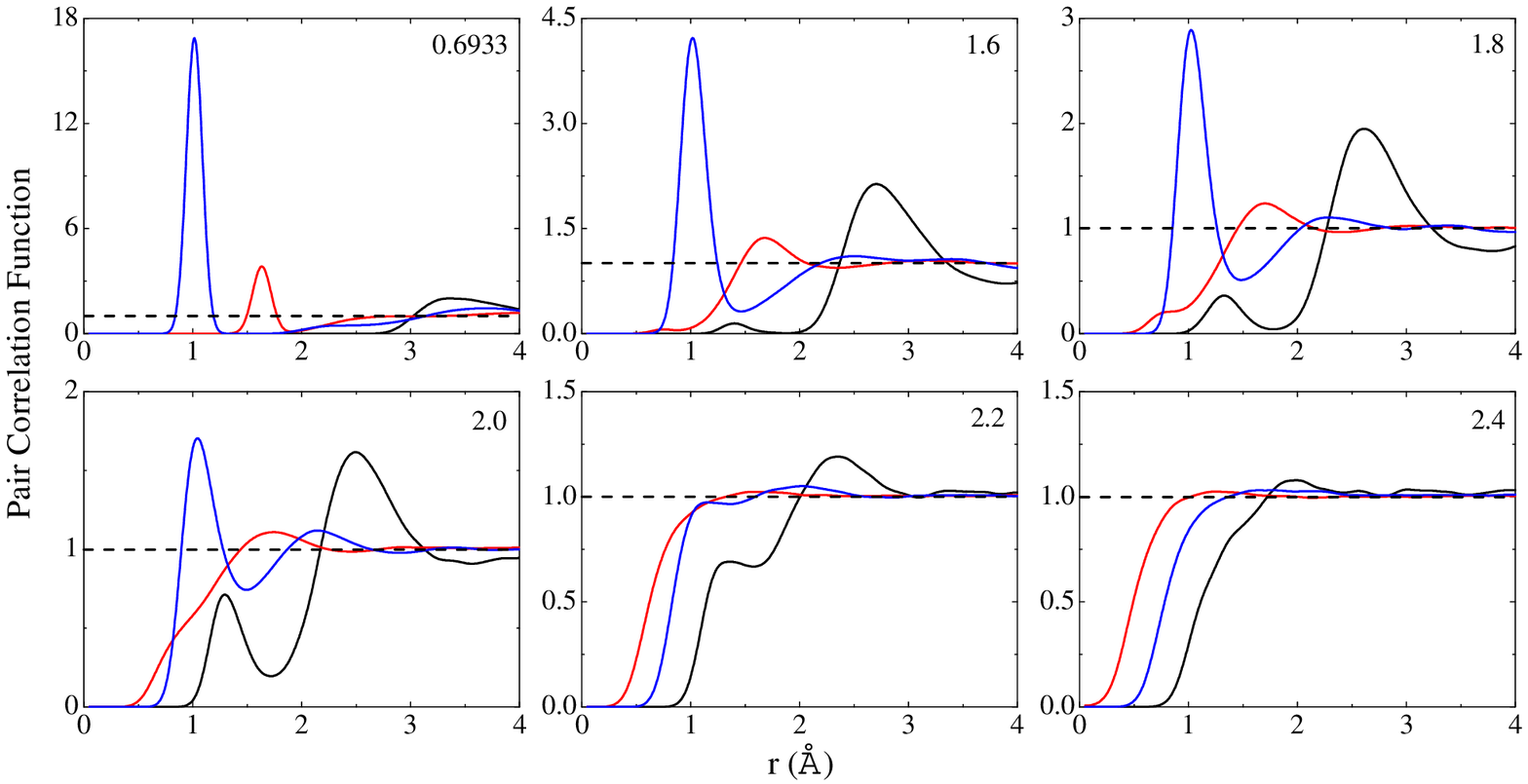}
\caption{Pair-correlation functions for N-N (black line), N-H (blue
line), H-H (red line) along the principal ammonia Hugoniot.
(a)$\rho=0.6933\,\textnormal{g/c\ensuremath{m^{3}}}$,
$T=230\,\textnormal{K}$;
(b)$\rho=1.6\,\textnormal{g/c\ensuremath{m^{3}}}$,
$T=3082\,\textnormal{K}$;
(c)$\rho=1.8\,\textnormal{g/c\ensuremath{m^{3}}}$,
$T=4174\,\textnormal{K}$;
(d)$\rho=2.0\,\textnormal{g/c\ensuremath{m^{3}}}$,
$T=6036\,\textnormal{K}$;
(e)$\rho=2.2\,\textnormal{g/c\ensuremath{m^{3}}}$,
$T=19180\,\textnormal{K}$;
(f)$\rho=2.4\,\textnormal{g/c\ensuremath{m^{3}}}$,
$T=54229\,\textnormal{K}$.}
\end{figure*}

\subsection{Dynamic optical and electronic properties}

\begin{figure}
\includegraphics[bb=0bp 0bp 568bp 402bp,clip,scale=0.5]{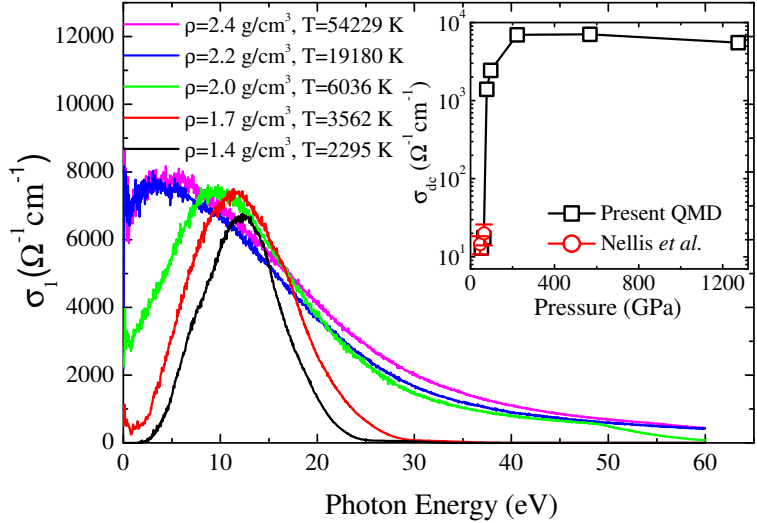}
\caption{The real part of electrical conductivity
$\sigma_{1}\left(\omega\right)$ along the principal Hugoniot. Data
have been averaged over 10 uncorrelated MD configurations. (Inset)
dc conductivity of our results and experimental data from Ref. 9 are
plotted along the principal Hugoniot. }
\end{figure}

Using the Kubo-Greenwood formula, we calculate the real part of
frequency-dependent conductivity $\sigma_{1}\left(\omega\right)$ for
points along the Hugoniot (with the initial condition
$\rho_{0}\mathtt{=}$0.6933 g/cm$^{3}$, $T_{0}$=230 K), as shown in
Fig. 5. Calculation of $\sigma_{1}\left(\omega\right)$ given by Eq.
(1) typically involves 1000 states, which insures adequate
convergence over the frequency range we considered. The maximum of
$\sigma_{1}\left(\omega\right)$ around 10 eV can be attributed to
the transitions to the lowest excited states. It is found that the
peak moves to lower frequency with increasing density and
temperature, and thus leads to a significant increase in dc
conductivity, which is defined as
$\sigma_{\textnormal{dc}}=\underset{\omega\rightarrow0}{lim}\sigma_{1}\left(\omega\right)$.
The significant variation of the dc conductivity along the principal
Hugoniot is highlighted in Fig. 5 as an inset. We first notice that
the dc conductivity becomes nonzero when approaching 50 GPa. For
pressures below 65 GPa, our calculated dc conductivity agrees well
with the attainable experimental data. After that, the dc
conductivity rises rapidly to a value larger than 1000
$\Omega^{-1}\textnormal{c\ensuremath{\textnormal{m}^{-1}}}$ for
pressures between 65 and 75 GPa. Based on the definition of
metallicity of the disordered system which has been discussed for
warm dense methane \cite{Li2011}, it could be concluded that
nonmetal-metal transition takes place in shocked ammonia. Then, a
plateau is reached near 220 GPa, which has been observed in some
other molecular fluids \cite{Boates2011,Chau2003}. Such a behavior
of the dc conductivity can be ascribed to the dissociation of the
molecules as discussed below.

\begin{figure}
\includegraphics[clip,scale=0.7]{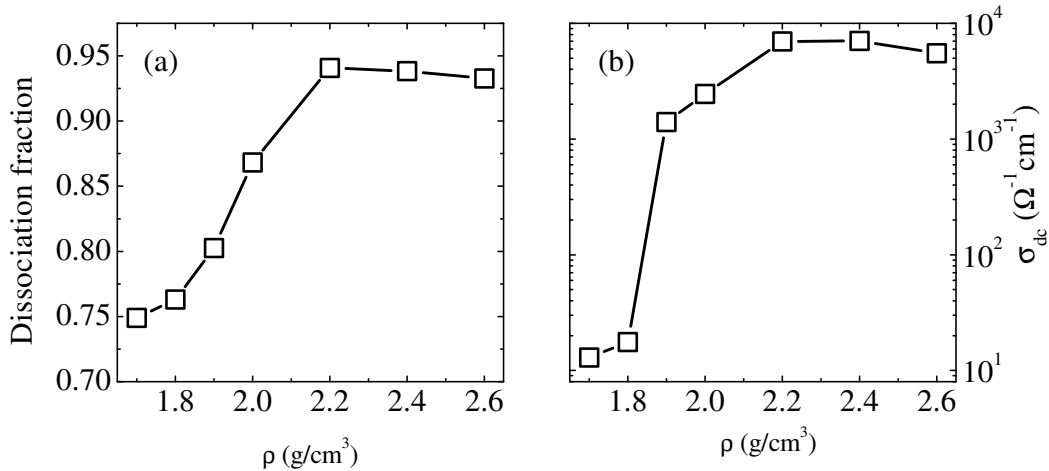}
\caption{(a) Dissociation fraction (b) Corresponding dc conductivity
$\sigma_{\textnormal{dc}}$ as a function of density along the
principal Hugoniot.}
\end{figure}

To clarify the nature of the fluid along the Hugoniot, we present in
Fig. 6(a) the variation of the dissociation fraction of ammonia
molecules as a function of density. Following the QMD simulations,
we identify ammonia molecules from the trajectories based on a
simple bond-length criteria. A cutoff radius of 1.25 $\textrm{\AA}$
is used to construct a sphere about each nitrogen atom and all
hydrogen atoms within this region are considered as bound to this
reference nitrogen atom. The number of ammonia molecules is counted
at each configuration and then averaged along the trajectory. As
seen from Fig. 6(a), the fluid has been partially dissociated at 1.7
g/cm$^{3}$. Whereas, at 2.2 g/cm$^{3}$ and above, the system is
fully dissociated. Similarly, from the conductivity curve shown in
Fig. 6(b), we can find that the variation of dc conductivity as a
function of density closely follows the variation of molecular
dissociation fraction. This intimate connection suggests that
dissociation has important influence on the electrical properties of
the system and results in the nonmetal-metal transition
consequently.

\begin{figure}
\includegraphics[bb=0bp 0bp 567bp 249bp,clip,scale=0.7]{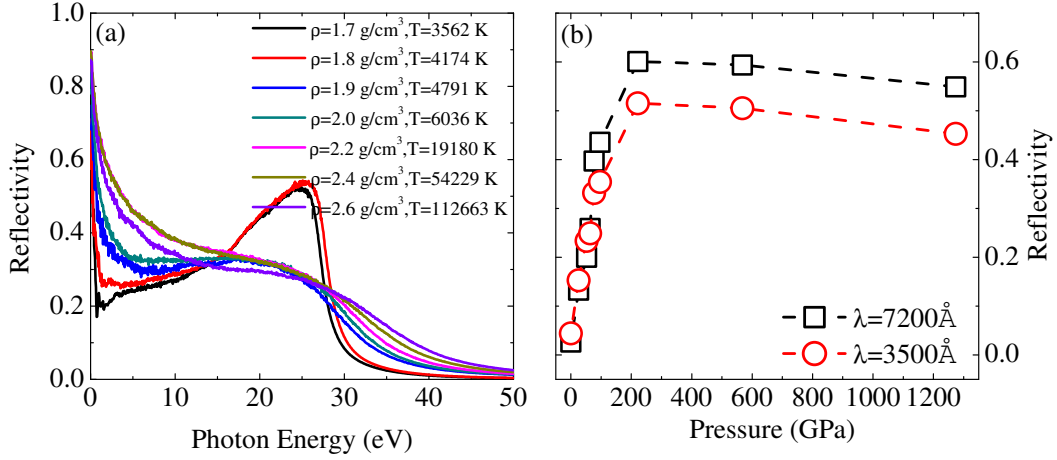}
\caption{(a) Frequency dependence of the reflectivity of ammonia
along the Hugoniot. (b) Reflectivity at wavelengths of 7200 (black
square) and 3500 $\textrm{\AA}$ (red circle) as a function of
pressure along the Hugoniot. }
\end{figure}

In dynamic shock compression experiments, optical reflectivity is
one of the readily observables. For ammonia, we show in Fig. 7(a)
the variation of the dynamic reflectivity $r\left(\omega\right)$
along the Hugoniot. With the increase of the density and
temperature, the shape of the curve changes abruptly at 1.9
g/cm$^{3}$, which is related with the high-pressure nonmetal-metal
transition. Correspondingly, the reflectivity at typical wavelengths
of 350 and 720 nm increases sharply around 75 GPa, as shown in Fig.
7(b). Furthermore, the increase in reflectivity mostly ceases and
reaches a plateau near 220 GPa where ammonia molecular are fully
dissociated. It can be seen that a measurable reflectivity increase
from 0.02 to 0.5-0.6 arises for the principal shocks.

\begin{figure}
\includegraphics[clip,scale=0.7]{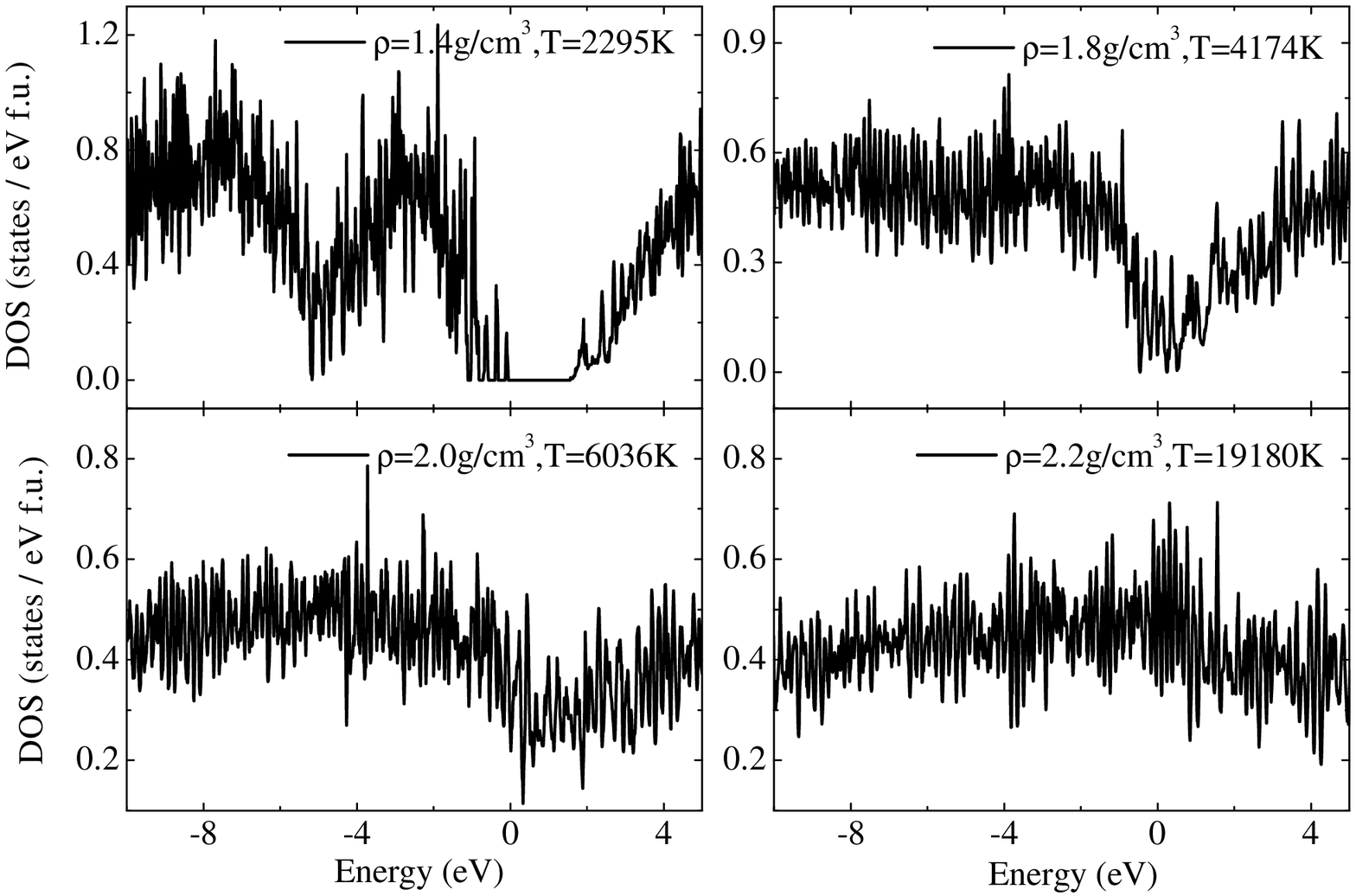}
\caption{Electronic density of states for liquid ammonia along the
principal Hugoniot:
(a)$\rho=1.4\,\textnormal{g/c\ensuremath{m^{3}}}$,
$T=2295\,\textnormal{K}$;
(b)$\rho=1.8\,\textnormal{g/c\ensuremath{m^{3}}}$,
$T=4174\,\textnormal{K}$;
(c)$\rho=2.0\,\textnormal{g/c\ensuremath{m^{3}}},\:
T=6036\,\textnormal{K}$;
(d)$\rho=2.2\,\textnormal{g/c\ensuremath{m^{3}}},\:
T=19180\,\textnormal{K}$. Data have been averaged over 10
uncorrelated MD configurations. The zero of the energy scale shows
the position of the Fermi level.}
\end{figure}

One way to characterize the behavior of nonmetal-metal transition is
the variation of the electron density of state (EDOS) along the
Hugoniot as shown in Fig. 8. It is found that a gap about 2.0 eV
appears at $\rho\mathtt{=}$1.4 g/cm$^{3}$ and  $T$=2295 K, with only
thermally activated electron transport occurring. With ammonia
molecules dissociate, the resulting atoms act as dopants and
progressively fill the dense fluid band gap, and eventually leads to
a metal-like behavior.

\section{CONCLUSIONS}

In conclusion, we have predicted the Hugoniot of ammonia up to 1.3
TPa, which is in good agreement with available shock experiments
with pressure up to 64 GPa. For comparison with future experiments,
we have also computed several Hugoniot curves with different initial
states. Three characterized segments are identified along the
Hugoniot. As the system dissociates and transforms into complex
mixture state, nonmetal-metal transition takes place, which is
determined through analyzing the electrical conductivity and optical
reflectivity. In addition, the isentropes of Uranus and Neptune pass
through the temperature-pressure conditions of the mixture regime
along the Hugoniot, while the superionic phase of ammonia exists far
below the isentropes. Therefore, we conclude that molecular hydrogen
could release into outer layers, which leads to more compact cores
in these planets.

\begin{acknowledgments}
This work was supported by NSFC under Grants No. 11205019 and No.
11275032, by the National Fundamental Security Research Program of
China, and by the Foundations for Development of Science and
Technology of China Academy of Engineering Physics under Grant No.
2009B0301037 and 2012B0102012.
\end{acknowledgments}

\end{document}